# Noise dependent Super Gaussian-Coherence based dual microphone Speech Enhancement for hearing aid application using smartphone

Nikhil Shankar, Gautam S Bhat, Chandan K A Reddy, Student Members, *IEEE*, Issa Panahi, Senior Member, *IEEE*

*Abstract*— **In this paper, the coherence between speech and noise signals is used to obtain a Speech Enhancement (SE) gain function, in combination with a Super Gaussian Joint Maximum a Posteriori (SGJMAP) single microphone SE gain function. The proposed SE method can be implemented on a smartphone that works as an assistive device to hearing aids. Although coherence SE gain function suppresses the background noise well, it distorts the speech. In contrary, SE using SGJMAP improves speech quality with additional musical noise, which we contain by using a post filter. The weighted union of these two gain functions strikes a balance between noise suppression and speech distortion. A "weighting" parameter is introduced in the derived gain function to allow the smartphone user to control the weighting factor based on different background noise and their comfort level of hearing. Objective and subjective measures of the proposed method show effective improvement in comparison to standard techniques considered in this paper for several noisy conditions at signal to noise ratio levels of -5 dB, 0 dB and 5 dB.**

## I. INTRODUCTION

Records by National Institute on Deafness and Other Communication Disorders (NIDCD) indicate that nearly 15% of adults (37million) aged 18 and over report some kind of hearing loss in the United States. Amongst the entire world population, 360 million people suffer from hearing loss. Over the past decade, researchers have developed many feasible solutions for hearing impaired in the form of Hearing Aid Devices (HADs) and Cochlear Implants (CI). However, the performance of the HADs degrade in the presence of different types of background noise and lacks the computational power, due to the design constraints and to handle obligatory signal processing algorithms [1-3]. Lately, HADs manufacturers are using a pen or a necklace as an external microphone to capture speech and transmit the signal and data by wire or wirelessly to HADs [4]. The expense of these existing auxiliary devices poses as a limitation. An alternative solution is the use of smartphone which can capture the noisy speech data using the two microphones, perform complex computations using the SE algorithm and transmit the enhanced speech to the HADs. There are many existing HADs applications which enhance the overall quality and intelligibility of the speech perceived by hearing impaired. Most of these applications use single microphone of the smartphone. Recent progressions include microphone array based SE techniques for better noise suppression. But, as the number of microphones increases, so does the cost and computational power. Thus, a dual microphone methodology was considered which provide a favorable solution for improving speech quality. In this work, we present a two microphone SE method that can be implemented on a smartphone as an application with a user interface.

Existing methods like SGJMAP single microphone SE [5] introduce musical noise due to half-wave rectification problem [6]. This can be solved by estimating the clean speech magnitude spectrum by minimizing a statistical error criterion, proposed by Ephraim and Malah [7, 8]. In [9], a computationally proficient alternative is proposed for SE methods in [7, 8]. In this method, super- Gaussian extension of the joint maximum a posteriori (JMAP) estimation rule is proposed to estimate the speech. By using the Super-Gaussian model of speech, mean squared error is minimized compared to Gaussian model [5].

Coherence based method dealing with coherent noise is appropriate for HADs and CI devices [10]. The theory behind these methods is that the speech from the two microphones is correlated, while the noise is uncorrelated with speech. Based on this, a gain function is defined to filter the noisy speech [10]. Using the coherence based function, noise is suppressed along with distortion in speech [3]. A weighted combination of the coherence gain function and SGJMAP SE gain resulted in better speech quality and intelligibility. The efficiency of this proposed method makes it computationally capable of implementing on smartphones to work seamlessly with HADs.

In this paper, we introduce a parameter called '*weighting'* factor in the proposed SE gain function that can be varied by the user to control the weighted combination, which in turn controls the amount of noise suppression and speech distortion. The parameter can be adjusted depending on the background noise. Various objective and subjective evaluations of the proposed method are carried out for the comparison of the method against the existing benchmark techniques considered.

## II. WEIGHTED COMBINATION OF COHERENCE-BASED AND SGJMAP GAIN FUNCTIONS

### A. *SGJMAP based speech enhancement*

In the SGJMAP [5] method, a non-Gaussianity property in spectral domain noise reduction framework is considered for the usage of super Gaussian speech model [11, 12]. Consider noisy speech $y(n)$, with clean speech $x(n)$ and noise $w(n)$,

*This work was supported by the National Institute of the Deafness and Other Communication Disorders (NIDCD) of the National Institutes of Health (NIH) under the grant number 5R01DC015430-02. The content is solely the responsibility of the authors and does not necessarily represent the official views of the NIH. The authors are with the Statistical Signal Processing Research Laboratory (SSPRL), Department of Electrical and Computer Engineering, The University of Texas at Dallas.

$$y(n) = x(n) + w(n) \quad (1)$$

The Discrete Fourier Transform (DFT) coefficient of $y(n)$ for frame $\lambda$ and $k^{th}$ frequency bin is given by,

$$Y_k(\lambda) = X_k(\lambda) + W_k(\lambda) \quad (2)$$

where $X$ and $W$ are the clean speech and noise DFT coefficients respectively. In polar coordinates, (2) can be written as,

$$R_k(\lambda)e^{j\theta_{Y_k}(\lambda)} = A_k(\lambda)e^{j\theta_{X_k}(\lambda)} + B_k(\lambda)e^{j\theta_{W_k}(\lambda)} \quad (3)$$

where $R_k(\lambda)$, $A_k(\lambda)$, $B_k(\lambda)$ are DFT amplitude of noisy speech, clean speech and noise respectively. $\theta_{Y_k}(\lambda)$, $\theta_{X_k}(\lambda)$, $\theta_{W_k}(\lambda)$ are the phases of noisy speech, clean speech and noise respectively. The purpose is to estimate clean speech magnitude spectrum $A_k(\lambda)$ and its phase spectrum $\theta_{X_k}(\lambda)$. $\lambda$ is dropped in further discussion for swiftness. The JMAP estimator of the amplitude and phase jointly maximize the probability of amplitude and phase spectra conditioned on the observed complex coefficient given by,

$$\hat{A}_k = \arg\max_{A_k} \frac{p(Y_k|A_k,\theta_{X_k})p(A_k,\theta_{X_k})}{p(Y_k)} \quad (4)$$

$$\hat{\theta}_{S_k} = \arg\max_{\theta_{X_k}} \frac{p(Y_k|A_k,\theta_{X_k})p(A_k,\theta_{X_k})}{p(Y_k)} \quad (5)$$

Using the super Gaussian speech model, spectral amplitude estimator allows the probability density function (PDF) of the speech spectral amplitude to be approximated by the function of two parameters $\mu$ and $v$. The super-Gaussian PDF [11] of the amplitude spectral coefficient with variance $\sigma_{S_k}$ is given by,

$$p(A_k) = \frac{\mu^{v+1}}{\Gamma(v+1)}\frac{A_k^v}{\sigma_{X_k}^{v+1}}\exp\left\{-\frac{\mu A_k}{\sigma_{X_k}}\right\} \quad (6)$$

where $\Gamma$ denotes the Gamma function.

Taking logarithm of (4), and differentiating with respect to $A_k$ gives,

$$\frac{d}{dA_k}\log(p(Y_k|A_k,\theta_{X_k})p(A_k,\theta_{X_k})) = \frac{-(Y_k^*-A_k e^{-j\theta_{X_k}})(-jA_k e^{j\theta_{X_k}})+(Y_k-A_k e^{j\theta_{X_k}})(jA_k e^{-j\theta_{X_k}})}{\hat{\sigma}_{W_k}^2} \quad (7)$$

Setting (7) to zero and substituting $Y_k = R_k e^{j\theta_{Y_k}}$ simplifies to

$$\frac{2R_k}{\hat{\sigma}_{W_k}^2} - \frac{2A_k}{\hat{\sigma}_{W_k}^2} + \frac{v}{A_k\beta} - \frac{\mu}{\hat{\sigma}_{X_k}} = 0 \quad (8)$$

On simplifying (8), the following quadratic equation is obtained,

$$A_k^2 + \frac{A_k}{2\hat{\sigma}_{X_k}}\left(\hat{\sigma}_{W_k}^2\mu - 2R_k\hat{\sigma}_{X_k}\right) - \frac{v\hat{\sigma}_{W_k}^2}{2} = 0 \quad (9)$$

Solving the above quadratic equation and writing in terms of $\hat{\xi}_k$ and $\hat{\gamma}_k$ yields

$$\hat{A}_k = \left(u + \sqrt{u^2 + \frac{v}{2\hat{\gamma}_k}}\right)R_k,\; u = \frac{1}{2} - \frac{\mu}{4\sqrt{\hat{\gamma}_k\hat{\xi}_k}} \quad (10)$$

where $\hat{\xi}_k = \frac{\hat{\sigma}_{X_k}^2}{\hat{\sigma}_{W_k}^2}$ is the *a priori* SNR and $\hat{\gamma}_k = \frac{R_k^2}{\hat{\sigma}_{W_k}^2}$ is the a posteriori SNR. $\hat{\sigma}_{W_k}^2$ is estimated using a voice activity detector (VAD) [13]. $\hat{\sigma}_{X_k}$ is the estimated instantaneous clean speech power spectral density. In [5], $v = 0.1$ and $\mu = 1.5$ is shown to give better results. The optimal phase spectrum and the noisy phase are assumed the same $\hat{\theta}_{X_k} = \theta_{Y_k}$.

The speech magnitude spectrum estimate is $\hat{A}_k = G_k R_k$ (11)

where $$G_k = \left[u + \sqrt{u^2 + \frac{v}{2\hat{\gamma}_k}}\right] \quad (12)$$

As in [14], it is considered that phase is perceptually insignificant. For the complete derivation of the gain function, we refer [5].

### *B. Coherence based gain function*

In this paper, we consider two microphones of Google Pixel placed apart (by about 13 cm) such that speech source and noise source are separated spatially and assumed to be at angles 0 and $\theta$ respectively [3], where $0^0 \le \theta \le 360^0$. The noisy speech is defined as,

$$\mathrm{y(n)} = x_j(n) + w_j(\mathrm{n}) \qquad (j = 1, 2) \quad (13)$$

where $j$ is the microphone index, $x_j(n)$ and $w_j(\mathrm{n})$ are speech and noise components respectively at each microphone. The Short Time Fourier Transform (STFT) of (13) is defined as,

$$Y_j(\omega, l) = X_j(\omega, l) + N_j(\omega, l) \qquad (j = 1, 2) \quad (14)$$

where $l$ is the frame index and $\omega = 2\pi n/N$, where $n = 0, 1, 2, \ldots, N-1$, $N$ is the number of FFT points. $\omega$ lies in the range of $[-\pi, \pi]$. The complex coherence function between the two signals is given by,

$$\Gamma_{y_1y_2}(\omega, l) = \frac{\Phi_{y_1y_2}(\omega,l)}{\sqrt{(\Phi_{y_1y_1}(\omega,l)\Phi_{y_2y_2}(\omega,l))}} \quad (15)$$

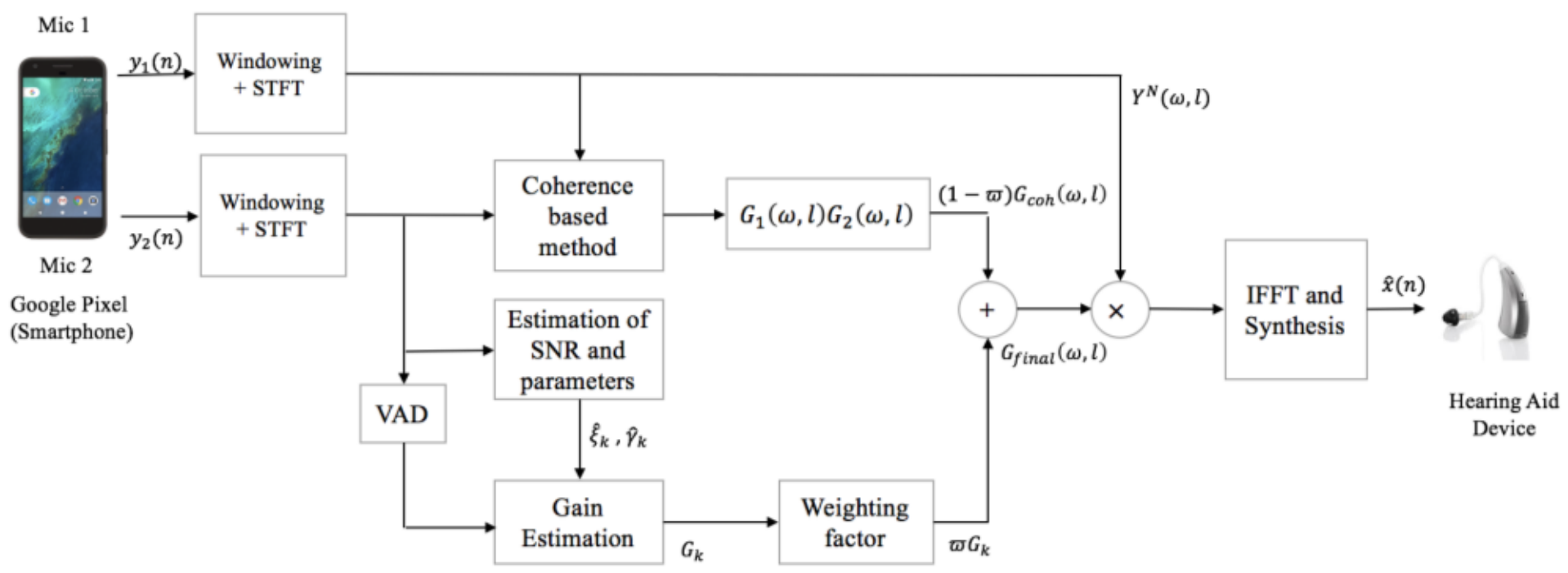

Fig. 1. Block diagram of proposed SE method

where $\Phi_{uv}(\omega, l)$ is the cross-power spectral density (CSD) defined as $\Phi_{uv}(\omega, l) = E[U(\omega, l)V^*(\omega, l)]$ and $\Phi_{uu}(\omega, l) = E[|U(\omega, l)|^2]$ is the power spectral density (PSD). According to [3], the noise and speech components are uncorrelated. Therefore, CSD of the microphone signals can be written as,

$$\Phi_{y_1y_2}(\omega, l) = \Phi_{x_1x_2}(\omega, l) + \Phi_{n_1n_2}(\omega, l) \tag{16}$$

Dividing both sides of (16) by $\sqrt{(\Phi_{y_1y_1}(\omega, l)\Phi_{y_2y_2}(\omega, l))}$ and neglecting $\omega$ and $l$ we get,

$$\Gamma_{y_1y_2} = \Gamma_{x_1x_2}\sqrt{\frac{\Phi_{x_1x_1}}{\Phi_{x_1x_1} + \Phi_{n_1n_1}}}\sqrt{\frac{\Phi_{x_2x_2}}{\Phi_{x_2x_2} + \Phi_{n_2n_2}}} + \Gamma_{n_1n_2}\sqrt{\frac{\Phi_{n_1n_1}}{\Phi_{x_1x_1} + \Phi_{n_1n_1}}}\sqrt{\frac{\Phi_{n_2n_2}}{\Phi_{x_2x_2} + \Phi_{n_2n_2}}} \tag{17}$$

True Signal to Noise Ratio (SNR) at the j$^{th}$ microphone is given by,

$$SNR_j = \frac{\Phi_{x_jx_j}}{\Phi_{n_jn_j}} \quad (j = 1, 2) \tag{18}$$

Considering the closeness of the two microphones, we can assume that $SNR_1 \approx SNR_2 \approx \widehat{SNR}$. It can be seen that coherence of speech is dominant at high SNR values and that of noise is dominant at low SNR values. From an approximation for the coherence value given in [10], (17) can be written as,

$$\hat{\Gamma}_{y_1y_2} = [\cos(\omega\tau) + j\,sin(\omega\tau)]\frac{\widehat{SNR}}{1 + \widehat{SNR}} + [\cos(\omega\tau\,cos\theta) + j\,sin(\omega\tau\,cos\theta)]\frac{1}{1 + \widehat{SNR}} \tag{19}$$

where $\tau = f_s(d/c)$, d is the microphone spacing, c is the speed of sound and $f_s$ is the sampling frequency. We make use of a suppression filter proposed in [10], where in 2 separate filters are used to suppress noise from certain range of $\theta$ values. For $\theta$ values around $90^0$, the suppression filter is,

$$G_1(\omega, l) = 1 - \left|\mathcal{R}\left[\hat{\Gamma}_{y_1y_2}(\omega, l)\right]\right|^{P(\omega)} \tag{20}$$

where $\mathcal{R}[.]$ is the real part and $P(\omega)$ is defined in two frequency bands as,

$$P(\omega) = \begin{cases}\alpha_{low}, & if\ |\omega| \le \frac{\pi}{8}\\ \alpha_{high}, & if\ |\omega| > \frac{\pi}{8}\end{cases} \tag{21}$$

where $\alpha_{low}$ and $\alpha_{high}$ are positive integer constants such that $\alpha_{low} > \alpha_{high} > 1$. When $90^0 < \theta \le 180^0$, the gain function becomes,

$$G_2(\omega, l) = \begin{cases}\mu, & if\ \Im\left(\hat{\Gamma}_{y_1y_2}(\omega, l)\right) < Q(\omega)\\ 1, & Otherwise\end{cases} \tag{22}$$

where $Q(\omega)$ is defined as,

$$Q(\omega) = \begin{cases}\beta_{low}, & if\ |\omega| \le \frac{\pi}{8}\\ \beta_{high}, & if\ |\omega| > \frac{\pi}{8}.\end{cases} \tag{23}$$

where $\beta_{low}$ and $\beta_{high}$ are negative constants such that $\beta_{low} > \beta_{high} > -1$. For further details on these gain functions, we refer to [10]. The final coherence based gain function is defined as,

$$G_{coh}(\omega, l) = G_1(\omega, l)G_2(\omega, l) \tag{24}$$

### C. *Weighted Combination of $G_k$ and $G_{coh}(\omega, l)$*

The block diagram of the proposed method is as shown in Figure 1. Windowing and STFT is applied on to the two microphone signals to convert them to frequency domain [3]. Though the coherence based gain function in (24) suppresses noise, the speech signal sounds somewhat mechanical. The quality of the speech can be retained by using $G_k$, which is the SGJMAP SE gain, but it introduces musical noise for background noise types such as babble or car noise which are non-stationary in nature. To bypass this limitation, we introduce a new gain function $G_{final}(\omega, l)$ given by,

$$G_{final}(\omega, l) = \varpi G_k + (1 - \varpi)G_{coh}(\omega, l) \tag{25}$$

where $\varpi$ is the weighting factor that helps the user to switch between noise suppression and speech distortion. At high values of $\varpi$, the final gain $G_{final}(\omega, l)$ results in good noise suppression and limited speech distortion.

## III. Experimental Evaluation

In this paper, Google Pixel with Android 7.1 Nougat operating system is considered as an assistive device for HA. The above device has an M4/T4 HA Compatibility rating and meets the requirements set by Federal Communications Commission (FCC). The noisy speech was considered at the sampling rate of 16 kHz and 20 ms frames with 50% overlap. The computational time for each frame is around 12 ms on Pixel. The parameter values are hard coded based on [10]. The values of $v$ and $\mu$ are set as 0.1 and 1.5 respectively for the test results but can be varied depending on the noisy environment. The range of $\varpi$ is from 0 to 1. At $\varpi = 0.7$ it is seen to provide better results. The evaluation of the proposed method is assessed using both objective and subjective measures.

### A. *Reference Objective Results*

Realistic recordings of machinery and babble noise are added to speech signals taken from IEEE corpus [15]. For the objective measure of quality of speech, we use Perceptual Evaluation of Speech Quality (PESQ) [16]. Coherence Speech Intelligibility Index (CSII) [17] is used to measure the intelligibility of speech. PESQ ranges between 0.5 and 4, with 4 being high speech quality. CSII ranges between 0 and 1, with 1 being high intelligibility. Figure 2 shows the plots of PESQ and CSII versus SNR for two noise types. In comparison with single and dual microphone SE methods such as log-MMSE and coherence based techniques respectively, the proposed method gives better values of PESQ and CSII as shown in Figure 2 for machinery and babble noise types. For Stationary noise types (Machinery), the PESQ and CSII values of the proposed method show significant improvement.

### B. *Subjective Test Results*

Along with Objective measures, we perform Mean Opinion Score (MOS) tests on 20 normal hearing both male and female subjects. They were presented with noisy speech and enhanced speech using the proposed, log-MMSE and coherence methods at different SNR levels of -5 dB, 0 dB, and 5 dB. The subjects had to choose a suitable $\varpi$ based on the level of comfort, but were also instructed regarding the standard value.

Each subject was instructed to score in the range 1 to 5 for the different audio files based on the following criteria: 5 being excellent speech quality and imperceptible level of distortion. 4 for good speech quality with perceptible level of distortion. 3 stood for fair speech quality with mediocre level of distortion. 2 for poor speech quality with lot of disturbances, causing uneven distortions. 1 having the least quality of speech and intolerable level of distortion. Subjective test results are shown in Figure 3, which illustrates the effectiveness of the proposed method in various background noise, simultaneously upholding the speech quality and intelligibility.

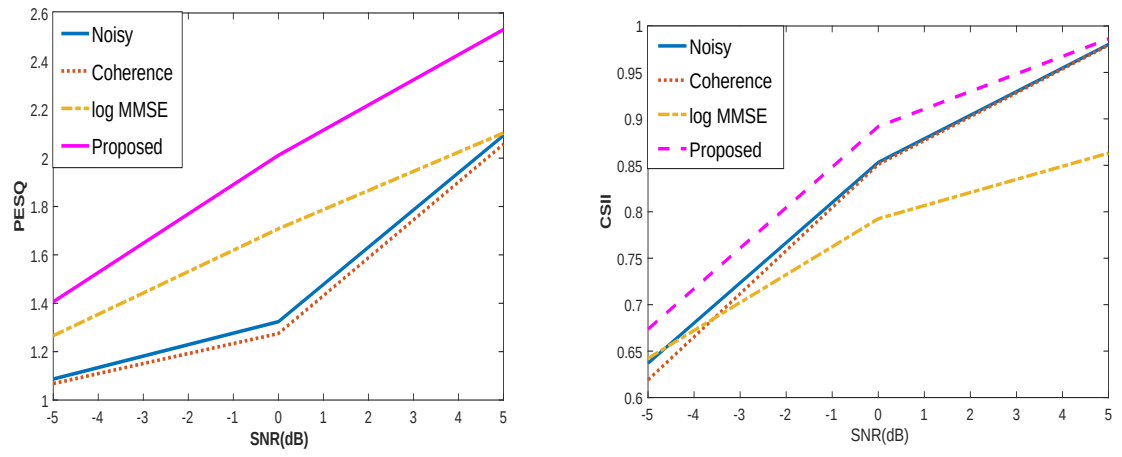

(a) Machinery noise

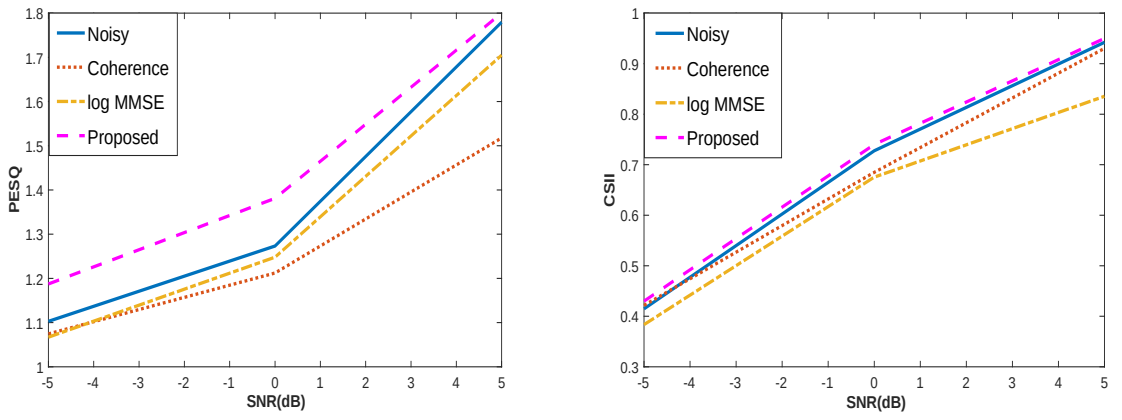

(b) Multi talker Babble noise

Fig.2. Objective measures of speech quality and intelligibility

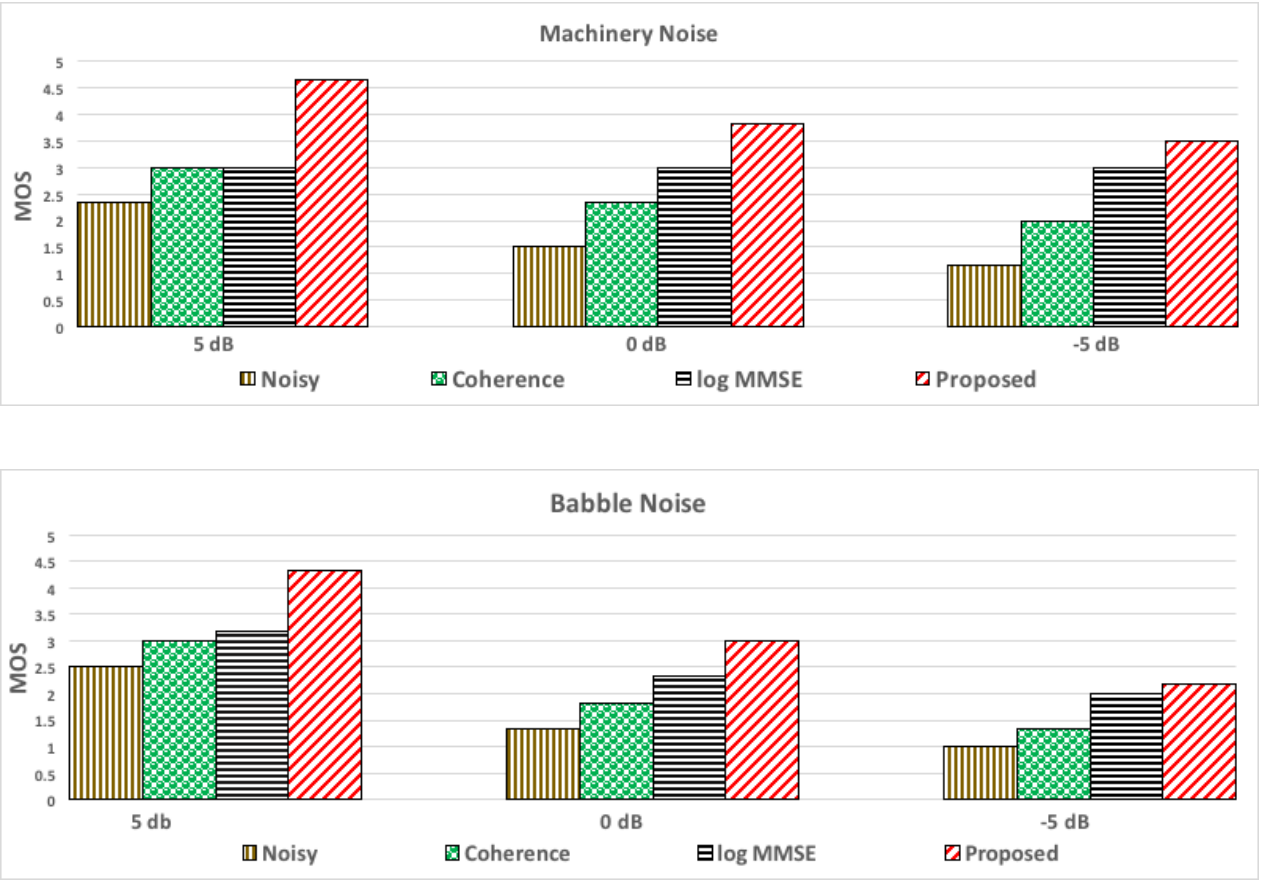

Fig. 3. Subjective test results

## IV. Conclusion

A Noise Dependent Super Gaussian - Coherence based dual microphone SE algorithm was developed with a weighting factor in the gain function. The obtained gain allows HAD user to control the amount of noise suppression and speech distortion. The proposed algorithm can run on a smartphone device in real time, which works as an assistive device for HA. The weighting parameter permits the smartphone user to control the amount of noise suppression (quality) and speech distortion (intelligibility). The objective and subjective evaluations verify the proposed method to be an apt option to use for hearing aid application in the real-world noisy environment.

## References

[1] Y-T. Kuo, T-J. Lin, W-H Chang, Y-T Li, C-W Liu and S-T Young, “Complexity-effective auditory compensation for digital hearing aids,” *IEEE Int. Symp on Circuits ad Systems (ISCAS)*, May 2008.

[2] Bhat, Gautam S., et al. "A real-time convolutional neural network based speech enhancement for hearing impaired listeners using smartphone." IEEE Access 7 (2019): 78421-78433.

[3] C. K. A. Reddy, Y. Hao, I. Panahi, “Two microphones spectral-coherence based speech enhancement for hearing aids using smartphone as an assistive device,” *IEEE Int. Conf. on Eng. In Medicne and Biology soc.*, Oct 2016.

[4] B. Edwards, “The future of Hearing Aid technology,” *Journal List, Trends Amplif*, v.11(1): 31-45, Mar 2007.

[5] Lotter, P. Vary, “Speech Enhancement by MAP Spectral Amplitude Estimation using a super-gaussian speech model,” *EURASIP Journal on Applied Sig. Process*, pp. 1110-1126, 2005.

[6] M. Berouti, M. Schwartz, and J. Makhoul, “Enhancement of speech corrupted by acoustic noise,” *Proc of IEEE Conf. on Acoustic SpeechSignal Processing*, pp. 208-211, Washington D.C, 1979.

[7] Y. Ephraim and D.Malah, “Speech enhancement using a minimum mean-square error short-time spectral amplitude estimator,” *IEEE Trans. Acoustics, Speech, and Signal Processing*, vol. 32, no. 6, pp. 1109–1121, 1984.

[8] Y. Ephraim and D.Malah, “Speech enhancement using a minimum mean-square error log-spectral amplitude estimator,” *IEEE Trans. Acoustics, Speech, and Signal Processing*, vol. 33, no. 2, pp. 443–445, 1985.

[9] Reddy, Chandan Karadagur Ananda, et al. "An individualized super-Gaussian single microphone speech enhancement for hearing aid users with smartphone as an assistive device." IEEE signal processing letters 24.11 (2017): 1601-1605.

[10] N. Yousefian and P. Loizou, “A Dual-Microphone Speech Enhancement algorithm based on the Coherence Function,” IEEE Trans. Audio, Speech, and Lang. Processing, vol. 20, no.2, pp. 599-609, Feb 2012.

[11] R. Martin, “Speech enhancement using MMSE short time spectral estimation with gamma distributed speech priors,” in *Proc. IEEE Int. Conf. Acoustics, Speech, Signal Processing (ICASSP ’02)*, vol. 1, pp. 253–256, Orlando, Fla, USA, May 2002.

[12] R. Martin and C. Breithaupt, “Speech enhancement in the DFT domain using Laplacian speech priors,” in *Proc. International Workshop on Acoustic Echo and Noise Control (IWAENC’03)*, pp. 87–90, Kyoto, Japan, September 2003.

[13] J. Sohn, N. S. Kim, and W. Sung, “A statistical model-based voice activity detection,” *IEEE Signal Processing Letters.*, vol. 6, no. 1, pp. 1–3, 1999.

[14] P. Vary, “Noise suppression by spectral magnitude estimation—mechanisms and theoretical limits,” *Signal Processing*, vol. 8, no. 4, pp. 387–400, 1985.

[15] “IEEE recommended practice for speech quality measurements,” IEEE Trans. Audio Electroacoust., vol. AE-19, no. 3, pp. 225–246, Sep. 1969.

[16] A. W. Rix, J. G. Beerends, M. P Hollier, A. P. Hekstra, “Perceptual evaluation of speech quality (PESQ) – a new method for speech quality assessment of telephone networks and codecs,” *IEEE Int. Conf. Acoust., Speech, Signal Processing (ICASSP)*, 2, pp. 749-752.

[17] P. Loizou, “Speech Enhancement: Theory and Practice”, Boca Raton, FL: CRC Press, 2007.